%% file: submission.tex
\documentclass[USenglish,oneside,twocolumn]{article}
\pdfoutput=1

\usepackage{times}
\usepackage{etex}
\usepackage[utf8]{inputenc}
\usepackage{amsmath}
\usepackage[hyphens]{url}
 
\DeclareMathOperator{\Enc}{Enc}

\DeclareMathOperator{\de}{deg}

\usepackage{subcaption}
\usepackage{pgfplots}

  \title{Open, privacy-preserving protocols for lawful surveillance}
  \author{
	Aaron Segal and Joan Feigenbaum \\
	Yale University \\
	New Haven, CT, USA\\
	\{aaron.segal,joan.feigenbaum\}@yale.edu\\
	\and
	Bryan Ford \\
	Swiss Federal Institute of Technology (EPFL) \\
	Lausanne, Switzerland\\
	bryan.ford@epfl.ch}
  \date{}

\begin{document}






\maketitle
 
  \begin{abstract}
{The question of how government agencies can
acquire actionable, useful information about legitimate but unknown targets 
without intruding upon the electronic activity of innocent parties is 
extremely important.  We address this question by providing experimental
evidence that actionable, useful information can indeed be obtained in a manner
that preserves the privacy of innocent parties and that holds government
agencies accountable.  In particular, we present practical, privacy-preserving
protocols for two operations that law-enforcement and intelligence agencies
have used effectively: {\it set intersection} and {\it contact chaining}.
Experiments with our protocols suggest
that privacy-preserving contact chaining can perform
a 3-hop privacy-preserving graph traversal producing 27,000 ciphertexts
in under two minutes.
These ciphertexts are usable in turn via privacy-preserving set
intersection to pinpoint potential unknown targets within a body of 150,000
total ciphertexts within 10 minutes,
without exposing personal information about non-targets.
}
  \end{abstract}



\maketitle

\input{intro}

\input{open}

\input{intersection}
\input{chaining}
\input{related}
\input{conc}


\bibliography{submission}
\bibliographystyle{plain}

\end{document}

%% file: intro.tex
\section{Introduction}\label{sec-introduction}
As networked devices become more available, more capable, and more ubiquitous
in everyday life, tension mounts between users' desire to safeguard their 
personal information and government agencies' desire to use that personal 
information in their pursuit of criminals and terrorists.  For example, 
the heated (and still unresolved) discussion about the Snowden 
revelations that started in 2013 is understood by many people
as an example of an unpleasant, stark choice: Citizens can either have
control over their personal information, or they can have
law-enforcement and intelligence agencies with the tools that they need to
keep the country safe. We regard this stark choice as a false dichotomy and 
assert that, by deploying appropriate security technology in the context of 
sound policy and the rule of law, we can have both user privacy and effective 
law enforcement and intelligence.

We draw a distinction between law-enforcement access to
third-party records held by businesses such as telephone companies
and encrypted information stored on an individual's personal devices;
our work focuses on the former, not the latter.
In particular, we believe that lawful searches of third-party records
can be made much more security- and privacy-preserving than they currently are
without compromising law-enforcement search capabilities.
We are {\em not} suggesting support for ``key-escrowed'' encryption
or general backdoors in personal devices,
as is the topic of the ongoing conflict between the FBI and Apple.

Adopting the approach to third-party records search
taken by Bandits~\cite{sff-foci2014},
we seek to design and implement protocols for {\it accountable surveillance}.
We require that government surveillance be conducted according to {\it open
processes}, as defined in Section~\ref{sec-open} below, and that the privacy
of {\it untargeted users} be protected.  We consider the surveillance goals
of {\it set intersection} and {\it contact chaining} and show that both can 
be achieved in a privacy-preserving, accountable fashion.

The utility of set-intersection protocols was demonstrated in the 
High Country Bandits case~\cite{anderson13cell}.
After obtaining cell-tower dumps -- sets of about 150,000 total users
whose cell phones had been in the vicinity of three banks at the times
that those banks were robbed -- the FBI intersected the sets and discovered 
that a single phone had been used at all of the relevant times in all of the 
relevant places.  They arrested the owner
of that phone and were able to prove that he was one of the robbers.
Although this FBI dragnet was effective in catching robbers, it also swept
in the cell-phone numbers of approximately 149,999 innocent bystanders.
Bandits~\cite{sff-foci2014} provide an accountable protocol for
set intersection that preserves the privacy of innocent bystanders.  Their
rudimentary implementation requires just under two hours on a test instance
with 150,000 total users.  In Section~\ref{sec-intersection} below, we provide
a more careful implementation that is faster by a factor of 10; in particular,
it runs for approximately 10.5 minutes on the test instance of 150,000 users.

In Section~\ref{sec-chaining}, we turn our attention to 
accountable-surveillance protocols for contact chaining.
The goal of contact chaining is to use the topology of a {\it communication 
graph} ({\it e.g.}, a phone-call graph, email graph, or social network) 
in order to identify associates (or ``contacts'') of lawfully targeted users~\cite{techdirt}. 
Government agencies can then investigate those associates to determine whether
they deserve further attention.\footnote{Note that contact chaining is not
tantamount to ``guilt by association.''  Rather, it is tantamount to
``suspicion by association,'' which is in fact a time-honored principal in
law enforcement.  When investigating a murder, rape, or other violent felony,
police departments first turn their attention to the associates of the victim.
Accountable-surveillance practices require that contacts who are suspected,
investigated, and found {\it not} to be involved in illegal activity be 
cleared of suspicion and that their personal information be deleted from 
agencies' databases.} It is useful to consider both direct contacts, 
{\it i.e.}, users who are neighbors in the communication graph, and
extended contacts, {\it i.e.}, users who are at distance $k$ in the 
communication graph, for an appropriate constant $k$.
In a phone-call graph, if Alice calls Bob, 
and Bob calls Charlie, then Alice and Bob are direct contacts (as are Bob and 
Charlie), but Alice and Charlie are extended contacts (more precisely,
contacts at distance 2).  Without accountability and security mechanisms to
limit an investigation's scope, contact chaining in a mass-communication
network can sweep in a huge number of untargeted users.
In Section~\ref{sec-chaining}, we provide an accountable contact-chaining
protocol that bounds the scope of the search, uses encryption to protect 
untargeted users, and is computationally efficient (in that its time
complexity and communication complexity are both linear in the size of the 
output).

At first blush, it may seem that a symposium on ``privacy-enhancing 
technologies'' is an odd place for results about ``accountable surveillance.''
No doubt some in the PETS community wish to prevent government agencies 
(as well as large corporations and other powerful entities) from conducting 
{\it any} surveillance whatsoever.  As explained in \cite{sff-foci2014}, a 
global communication system entirely free of surveillance may be appealing in
the abstract, but it is not a very useful goal in practice.  Law enforcement
and intelligence agencies have always been and will continue to be active 
on the Internet and in all national- and global-scale communication systems.
The challenge for the technical community is to build systems that enable
government agencies to collect relevant data that they are legally authorized
to collect, to be held accountable to the citizens they serve, and to respect 
the privacy of innocent users.

%% file: open.tex
\section{The Openness Principle in Lawful Surveillance}\label{sec-open}
In this section, we review the openness principle put forth by
Bandits~\cite{sff-foci2014}.  Readers familiar with \cite{sff-foci2014} 
may skip to the next section.

Necessary to any meaningful discussion of ``accountable surveillance'' is an
established foundation of rule of law and democratic processes that subject
the laws to evaluation, debate, and revision.  Bulk surveillance must follow
{\it open processes}, {\it i.e.}, unclassified procedures laid out in public 
laws that all citizens have the right to read, to understand, and to challenge
through the political process.  Processes that are not open, public, and
unclassified in this sense are referred to as {\it secret processes}.  
Although government agencies need not always disclose all of the details of a 
particular investigation, they do need to follow the open processes 
established for all bulk surveillance.

More precisely, Bandits~\cite{sff-foci2014} draw a distinction 
between two classes of communication-system users.  {\it Targeted users} are 
those who are under suspicion and are targets of properly authorized warrants.
All others are {\it untargeted users}; they are the vast majority of all 
users of a general-purpose, mass-communication system. Bandits~\cite{sff-foci2014}
posit that the following {\it Openness Principle} should govern all surveillance
activity in a democratic society:
\begin{enumerate}
\item[I]
Any surveillance or law-enforcement process
that obtains or uses private information
about untargeted users shall be
an open, public, unclassified process.
\item[II]
Any secret surveillance or law-enforcement processes shall use only:
\begin{enumerate}
\item[(a)] public information, and
\item[(b)] private information about targeted users obtained under 
authorized warrants via open surveillance processes.
\end{enumerate}
\end{enumerate}
Bandits interpret this principle as a requirement that an open
``privacy firewall'' be placed between government agencies and citizens'
private information in a mass-communication network.  Processes that move
untargeted users' private information through the firewall must be open
processes.  The targeted class contains both {\it known} users, {\it i.e.},
those for whom the government has a name, address, phone number, email address,
or other piece of personally identifying information, and {\it unknown} users.
It is not, as it may seem on the surface, oxymoronic to call a user both
``targeted'' and ``unknown,'' because ambient information may justify the
targeting of an individual without identifying him or her in any standard sense
of ``identify.''  For example, a government agency may obtain a ``John Doe
warrant''~\cite{bieber-penn2002} to investigate users who were present in
locations $L_i$ at times $T_i$, for $1\leq i\leq k$, without being able to 
identify those users, because relevant events occurred at those locations at
those times. Bandits~\cite{sff-foci2014} show how an 
accountable-surveillance protocol can be used to obtain a large set of
{\it encrypted} data about both targeted and untargeted users, feed it into a 
cryptographic protocol that winnows it down to the records of users targeted by
the John Doe warrant, and decrypt {\it only} those records.  Thus, targeted
unknown users can be identified ({\it i.e.}, turned into targeted, known users)
without government agencies' identifying any untargeted users whose 
encrypted records are touched by the surveillance process.

The essence of the openness principle is that, by using
appropriate security technology, government agencies can make their 
data-collection {\it processes} fully public without revealing sensitive
{\it content} of a specific investigation.  For a more detailed explanation,
see \cite[Section 2]{sff-foci2014}.

%% file: intersection.tex
\section{Lawful Intersection Warrants}\label{sec-intersection}
In this section, we present an improved implementation of the lawful
set-intersection protocol of~\cite{sff-foci2014}. 
\subsection{Intersections and Privacy}

As described in Section~\ref{sec-introduction} and in~\cite{sff-foci2014}, the FBI used
set intersection to search for phone numbers that appeared in three sets of
cell-tower records. This procedure did not follow the the openness principle.
The FBI did not distinguish between targeted and untargeted users when
collecting the data. It only arrested the user in the intersection of the sets
it collected, as far as we know, but as there was no established accountability
procedure, we do not know whether the FBI retained the phone numbers of
untargeted users collected during the search, whether it subjected any of those
other phone numbers to additional investigation, whether it shared the data sets
of 150,000 total users with other government agencies, \emph{etc}. 

Bandits~\cite{sff-foci2014} presented a private set intersection
protocol, based on the ElGamal~\cite{elgamal} and Pohlig-Hellman~\cite{pohlighellman}
cryptosystems, to address these specific concerns. We present a summary of that protocol
in Section~\ref{sec-isectproto}. This protocol reveals \emph{only} the identities
of users in the intersection of the sets under consideration, leaving the other,
untargeted users' identities hidden by encryption. To provide accountability and
oversight, the protocol requires multiple government agencies to participate. This
provides a division of authority. No one agency can uncover users' identities
without other agencies being able to collect records about how often intersection
warrants are used and set restrictions on how many users the agencies can identify
under a single warrant.

The lawful contact chaining protocol we present in Section~\ref{sec-chaining} will
produce large, encrypted sets of user identities. These sets can then be used as
inputs into the lawful set intersection protocol, to reveal the small, targeted
group of identities that appear in multiple sets (whether those sets come from cell
tower dumps as in the High Country Bandits case, contact chaining, or some other
legally obtained source of information). 

\subsection{Lawful Intersection Protocol}\label{sec-isectproto}

The participants in this
protocol are government agencies. Before executing the protocol, the 
agencies must agree on which sets of encrypted user data they wish to 
intersect. These sets are then retrieved from a repository, which stores 
only ciphertexts encrypted with the ElGamal public keys of all agencies.

Each agency's input is its ElGamal private key and a set of this encrypted
data. For each execution of the protocol, the agencies also generate temporary 
Pohlig-Hellman keys, which they securely delete after the protocol is complete. 
The protocol works by converting ciphertexts in the probabilistic ElGamal 
cryptosystem to ciphertexts in the deterministic Pohlig-Hellman 
cryptosystem without revealing information about the encrypted data in the 
process. Because Pohlig-Hellman is deterministic, two identical user 
identifiers will have identical Pohlig-Hellman encryptions.

These two cryptosystems, ElGamal and Pohlig-Hellman, allow this 
conversion to take place because they are \emph{mutually commutative} with 
each other. That is, a ciphertext encrypted under some combination of 
multiple ElGamal encryption keys, multiple Pohlig-Hellman encryption keys, 
or a mixture of the two types of keys can be decrypted only by the 
corresponding set of decryption keys \emph{in any order}. ElGamal and 
Pohlig-Hellman are randomized and deterministic, respectively, and they 
satisfy the mutual commutativity requirement.

During the protocol, each agency in turn runs the ElGamal decryption 
algorithm on the ciphertexts using its private key, then runs the 
Pohlig-Hellman encryption algorithm using its temporary Pohlig-Hellman key. 
Because the ciphertexts are also encrypted under the keys of the other 
agencies, the agency does not learn anything about the underlying user 
data during this process. The agency then passes the altered ciphertexts on 
to the next agency, which does the same with its keys. At the conclusion of 
this process, the agencies have converted the ciphertexts from randomized 
ElGamal encryption to determinstic Pohlig-Hellman encryption, without 
ever revealing the plaintext data.

The agencies can then directly compare Pohlig-Hellman ciphertexts to each 
other to determine which appear in the intersection of all sets of data. 
If the intersection is much larger than expected, any one of the agencies can 
delete its Pohlig-Hellman decryption key to prevent any user data from being 
revealed until a more narrowly scoped warrant can be agreed upon. Otherwise, 
the agencies finish the protocol by using their Pohlig-Hellman decryption keys 
to reveal only the ciphertexts in the intersection of all sets.

\subsection{Improved Implementation of Lawful Intersection}\label{sec-isectimpl}

In \cite{sff-foci2014}, the authors presented a Java implementation of the 
lawful set-intersection protocol. It ran on three PlanetLab computers 
representing the participating government agencies. 
Although the servers split the data sets between them, the implementation 
handled each set in a sequential manner, decrypting and encrypting 
ciphertexts one by one.

In an experiment with 150,000 users - the same number of users as the FBI
examined in the High Country Bandits case - that implementation took
approximately two hours to run to completeion. As the authors argued an FBI
investigation is likely to take days to lead to an arrest, even without the
use of this privacy-preserving technology. Therefore, a two-hour running time
may not be a major obstacle in this context.

But because we offer a new use of the lawful intersection protocol, we
also offer a speedier implementation of it. Our improved implementation
takes advantage of parallel processing and more advanced computational 
hardware, thus showing that lawful set-intersection can be performed 
much more quickly than originally described in~\cite{sff-foci2014}. 

In our upgraded implementation, the agencies decrypt and encrypt 
multiple ciphertexts in parallel, rather than operating on them one by one. 
We use eight compute threads for each server. Instead of PlanetLab computers, 
which vary in speed and reliability, we use three separate hosts on a private 
cloud-computer system running on Open Stack with a Ceph storage backend.

\begin{figure}
\centering
\resizebox{\linewidth}{!}{\begin{tikzpicture}
\begin{loglogaxis}[
	title={Parallel vs. Sequential Running Time},
	xlabel={Total Set Size [Ciphertexts]},
	ylabel={Running Time [min]},
	xmin=10, xmax=200000,
	ymin=0.01, ymax=1000,
	legend pos=north west,
	ymajorgrids=true,
	xmajorgrids=true,
	grid style=dashed,
]

\addplot[
	color=blue,
	mark=o,
	]
	table [x=Items, y=Parallel, col sep=comma] {intersection.csv};
	
\addplot[
	color=red,
	mark=square,
	]
	table [x=Items, y=Sequential, col sep=comma] {intersection.csv};

\legend{New prototype (8 threads), Old prototype (1 thread)}
\end{loglogaxis}
\end{tikzpicture}
}
\captionsetup{justification=centering}
\caption{Comparison of Lawful Intersection Performance}
\label{fig:isectgraph}
\end{figure}
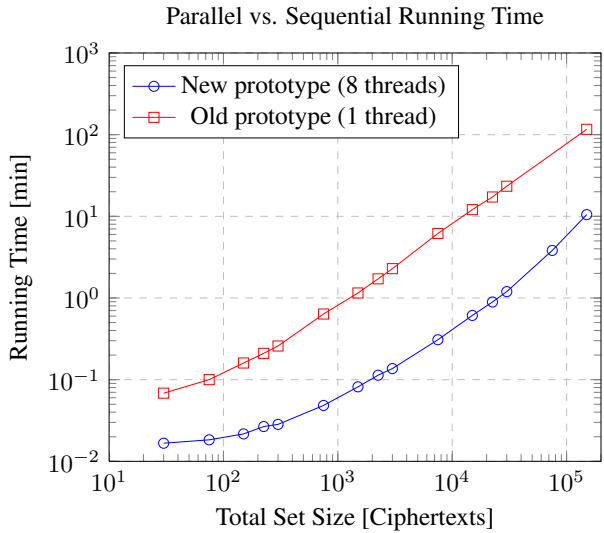

Compared with the orginal, sequential version of the protocol, our version 
requires only about 10\% as much time to run to completion. In the 
largest test case, which contains a total of 
150,000 ciphertexts (50,000 per server), our implementation takes 10.5 minutes, 
compared with 116.2 minutes for the orginal one. A full comparison of our 
results with the those of~\cite{sff-foci2014} is presented in 
Figure~\ref{fig:isectgraph}.

%% file: chaining.tex
\section{Lawful Contact Chaining}\label{sec-chaining}
Contact chaining~\cite{techdirt} is a form of government surveillance with which it is deceptively easy to expose many innocent, untargeted users to government scrutiny. The goal of contact chaining is to use information about social connections between identities, such as records of phone calls between one number and another, to identify members of a criminal organization or terrorist group. Starting with one or more suspects whose identities are known, the government aims to consider contacts of those suspects. These can be \emph{direct contacts}, such as two people who spoke on the phone, or \emph{extended contacts}, such as two people connected by a chain of two or more phone calls. If Alice calls Bob, and Bob calls Charlie, then Alice and Bob are direct contacts (as are Bob and Charlie), but Alice and Charlie are extended contacts. We may also say that Alice and Charlie are at distance 2 in the communication graph (because the smallest number of phone calls that connect Alice to Charlie is 2).

Without mechanisms to preserve privacy, a contact chaining search can collect a surprisingly large group of users' information. For example, if the average cell phone user contacts 30 individuals within the period of the investigation, a contact chaining search out to distance 3 would capture 27,000 users on average - or many more if a heavy phone user is swept up by the search. With such a large group, it is assured that the vast majority of contacts will not be the targeted collaborators of the primary suspect in the investigation. This is a large and unnecessary intrusion of privacy. These untargeted users may nevertheless face unwarranted government scrutiny, intrusive investigation, or a risk that their sensitive communications histories may be leaked accidentally.

Despite this risk, we recognize the potential law-enforcement value of information about social connections between targeted invidivuals. Therefore, we propose a \emph{lawful contact chaining protocol}. This protocol permits multiple government agencies working together to provide oversight and accountability, as advocated in \cite{sff-foci2014}. Our protocol focuses on the case in which the government seeks information from multiple telecommunications providers about the communication graph formed by phone calls and text messages. Using this protocol, the agencies can retrieve an encrypted set of user data from multiple telecoms, each of which holds only part of a larger communication graph. This encrypted set contains the identities of users within a certain distance of a target, but the identities cannot be decrypted unless the agencies cooperate. Under the lawful processes we propose, this cooperation would take the form of an intersecton with other encrypted sets of data, using the protocol from Section~\ref{sec-intersection}. These sets can come from privacy-preserving contact chaining, from cell tower dumps, or from other sources of information about suspects. Importantly, while any set may contain encrypted data about many untargeted users, very few users will appear in \emph{all} the sets, and those few users will be suitable targets for further lawful investigation.

The same principles of oversight and accountability provided by multiple government agencies can apply to contact chaining searches in other types of communication graphs, such as the social network graph of Twitter or Facebook. These cases do not require our protocol, however, since if one provider knows the entire communication graph, it can compute the output of the protocol without any interactivity needed.

\subsection{Protocols For Privacy-Preserving Contact Chaining}

\label{sec-proto}


\subsubsection{Inputs and Parties to the Protocol}

There are two types of parties in this protocol: Telecommunications companies (telecoms) and government agencies interested in performing lawful contact-chaining (agencies). The protocol computes a function of all parties' data.

The telecoms jointly hold an undirected communication graph $G=(V,E)$. Each telecom knows only a subset of the edges $E$. $V$ contains vertices labeled with the phone numbers they represent, and $E$ contains an edge between $a$ and $b$ if and only if phone number $a$ has contacted phone number $b$ or vice versa within some window of time. Each phone number $v$ is served by exactly one telecom. We assume telecoms know which telecom serves which phone number. Each telecom keeps records of all phone calls made by phones they serve, including calls made to phone numbers served by other telecoms. The subgraph known by telecom $T$ is $G_T=(V, E_T)$ where $E_T$ is the set of edges $(a, b)$ such that $a$ or $b$ is a phone number served by $T$. Henceforth, for any phone number $a$, let $T(a)$ be the telecom that serves $a$.

The agencies must each hold a copy of a \emph{warrant} in order to perform this protocol. A warrant is a triplet ($x$, $k$, $d$). $x$ is a target phone number. We assume, since $x$ belongs to a user targeted by the agencies, that they also know which telecom serves $x$. $k$ is a (small) distance from $x$, the distance at which the agencies wish to consider chained contacts. For example, if $k=2$, then the agencies only wish to consider users at most 2 phone calls away from their person (or phone number) of interest $x$. Choosing a small limit is important to limiting the scope of the investigation. However, many users' information might still be captured if some phone numbers have very many contacts. Suppose the target $x$ calls the most popular pizza place in town. Now everyone else who has recently called that pizza place is at a distance 2 to $x$.

We can assume that business phone numbers have many more contacts than personal phone numbers do. In most cases, knowing that two individuals have contacted the same business phone number does not indicate that those individuals have a personal relationship. Therefore, the warrant also includes $d$, an upper bound on the degree of users the agencies are willing to ``chain'' through. If a phone number has more than $d$ contacts, then the agencies do not consider paths to other users through that phone number in their search. The agencies disregard $d$ for the initial target $x$, however. The high-degree users themselves will also be present in the agencies' outputs.

This provides a reasonable limit to the scope of the investigation and hides what are very likely to be untargeted users from the government. In the uncommon scenario where a business number with many contacts also functions as a front or hub for a criminal organization to be revealed, the government is still able to conduct further investigation on it, perhaps even beginning a new contact-chaining search with that number as the initial target.

We specify the protocol in full in Sections~\ref{sec-proto1} and \ref{sec-proto2}.

\subsubsection{Security Assumptions}

We make a few assumptions about existing cryptographic infrastructure. All telecoms and agencies must have a public encryption key known to all other parties to the protocol and a private decryption key. For the purpose of interoperability with lawful intersection, agencies' keys must be for a commutative cryptosystem (i.e. ElGamal). The parties must also each have private signing keys and public verification keys.

In the protocol below, we refer to ``the agencies'' sending messages to one or more telecoms. Exactly which agency transmits messages to the telecoms is not important to our protocol, but a telecom will disregard any message not accompanied by signatures from every agency. One simple topology is for a single agency to handle all direct communication with telecoms and with other agencies, forwarding reponses from the telecoms on to the other agencies and signatures on agency messages to the telecoms.

Our protocol preserves the privacy of untargeted users as long as all parties execute the protocol in an honest-but-curious manner, all of the government agencies do not collude together, and no telecoms collude with government agencies. A colluding group containing all agencies would be equivalent to the current situation n which the government does not provide meaningful accountability of its own surveillance activities; what we propose is a replacement for this situation, but it does require the government to follow its own laws, once set. A telecom colluding with a government agency would amount to sending that agency free information about its users, or submitting incorrect information to the protocol. But telecoms have no business purpose to deviate from the protocol and risk legal action. In practice, existing legal tools allow law enforcement agencies to gather information about the phone history of a suspect with a valid warrant, but such information cannot generally be used for further contact chaining.

In Section~\ref{sec-future}, we discuss potental ways in which our honest-but-curious assumption might be relaxed.

\subsubsection{Desired Outputs and Privacy Properties}

The goal of the protocol is for the agencies to obtain a set of ciphertexts, each of which is the encryption of a phone number $v$ such that the distance in the communication graph from $v$ to the targeted phone number $x$ is at most $k$. The set should not contain encryptions of numbers $v$ such that each path from $x$ to $v$ of length at most $k$ contains an intermediate vertex of degree greater than $d$. Here the ``intermediate'' vertices in a path are all vertices except the endpoints $x$ and $v$.

Every phone number in this set must be encrypted with each of the agencies' public ElGamal keys. The agencies should all have the same output.

The telecoms should not learn the agency's output. Instead, each telecom's output should contain only a list of which of the phone numbers it serves were sent to the government agencies. This allows the telecoms to play an additional accountability role in this protocol. The government may have an interest in keeping the telecoms from knowing which of their clients were surveilled; we discuss this in section~\ref{sec:oblivious}.

With the encryptions of these phone numbers, the agencies can then act as appropriate to further investigate them. In particular, the encrypted set of phone numbers can be used as an input into a lawful set intersection protocol.

Below, we present two versions of our protocol. In the first version, the agencies and telecoms learn some additional information. Specifically, the agencies learn the provider of each phone number in the encrypted set, and the distance from $x$ of each encrypted phone number. Each telecom learns which of the phone numbers it serves appear in the agencies' output, as well as the distance of each of those phone numbers from the target phone number $x$.

In section \ref{sec-proto2}, we will present a second version of the protocol in which the agency \emph{does not} learn which telecom serves which encrypted phone number.

As long as our security assumptions for this protocol hold, the agencies collectively learn \emph{no} information about the edge set $E$ except what is implied by the output. Furthermore, the agencies cannot learn any of the phone numbers that appear in encrypted form in the output (unless implied by the size of the encrypted output and the leaked service information), nor can agencies cause a phone number not within distance $k$ of target $x$ to appear in the output, even in encrypted form.

\subsubsection{Ownership-Revealing Lawful Contact-Chaining Protocol}

\label{sec-proto1}

The protocol below amounts to a distributed breadth-first search of the communication graph run by the agencies making queries of the telecoms. However, all messages the agencies receive from the telecoms will be encrypted. They will know which message come from which telecoms, and will therefore know which telecoms serve which ciphertexts.

Let $\Enc_T(m)$ be the encryption of message $m$ under telecom $T$'s public key. Call such an encryption a \emph{telecom ciphertext}. Let $\Enc_\mathcal{A}(m)$ be the encryption of $m$ under the public keys of all agencies, and call such an encryption an \emph{agency ciphertext}.

To manage the breadth-first search, the agencies (or at least the investigating agency) will maintain a queue $\mathbf{Q}$, containing vertices yet to explore. $\mathbf{Q}$ contains tuples for unexplored vertices $a$ of the form $(\Enc_{T(a)}(a), T(a), j)$. These tuples contain the telecom ciphertext for $a$, a record of which telecom owns $a$, and an integer $j$ indicating the remaining distance out to which neighbors can be chained from $a$.

The agencies will represent their output in the form of a list $\mathbf{C}$, containing agency ciphertexts. Each telecom $T$ will represent its output in the form of a list $\mathbf{L}_T$, listing plaintext users served by that telecom whose information the agencies requested.

The protocol is as follows:

\begin{enumerate}

\item The agencies start by agreeing upon a warrant $(x, k, d)$, where $x$ is the target phone number, $k$ is a maximum distance, and $d$ is an upper limit on the degree of vertices to chain through. They encrypt $x$ under the public key of $T(x)$.

\item The agencies initialize a queue $\mathbf{Q}$. Initially, $\mathbf{Q}$ contains only the triple $(\Enc_{T(x)}(x), T(x), k)$.

\item The agencies initialize the output list $\mathbf{C}$ to be empty.

\item Each telecom $T$ initializes its output list $\mathbf{L}_T$ to be empty.

\item While $\mathbf{Q}$ is not empty, do the following:

\begin{enumerate}

\item \label{proto1:top-of-loop} The agencies dequeue $(\Enc_{T(a)}(a), T(a), j)$ from $\mathbf{Q}$. They send the pair ($\Enc_{T(a)}(a), j)$ to $T(a)$.

\item $a$'s provider, $T(a)$, decrypts $a$ from its telecom ciphertext. It adds $a$ to $\mathbf{L}_T$.

\item \label{proto1:first-send} $T(a)$ encrypts $a$ under the agencies' public keys, and sends $\Enc_\mathcal{A}(a)$ to the agencies.

\item If $j=0$, $T(a)$ is done. Go to step \ref{proto1:receive}.

\item Otherwise, $T(a)$ encrypts each neighbor $b$ of $a$ under the public key of $T(b)$, creating a telecom ciphertext for $b$.

\item \label{proto1:second-send} $T(a)$ sends the number of ciphertexts generated this way, $\de(a)$, as well as all telecom ciphertexts generated in the previous step, to the agencies. $T(a)$ sends the ciphertexts in the form of pairs $(\Enc_{T(b)}(b), T(b))$.

\item \label{proto1:receive} The agencies add $\Enc_\mathcal{A}(a)$ to $\mathbf{C}$.

\item If $\de(a)>d$ and $j\neq k$ (i.e. $a\neq x$), the agencies discard all telecom ciphertexts received for $a$'s neighbors (i.e., agencies refuse to sign these ciphertexts in future steps of the protocol, and do not send them on to the telecoms).

\item Otherwise, for each telecom ciphertext received, the agencies add $(\Enc_{T(b)}(b), T(b), j-1)$ to $\mathbf{Q}$.

\end{enumerate}

\item The agencies' final output is the list $\mathbf{C}$. Each telecom $T$'s final output is $\mathbf{L}_T$.

\end{enumerate}

For the sake of efficiency, it is worth noting that the inner loop can be executed many times in parallel, up to the point of completely emptying $\mathbf{Q}$ at the beginning of the loop. Many messages to the same telecom can also be batched and sent together, thereby reducing the number of signing and verifying operations so that they depend only on $k$ and not on the size of the input or output.

\subsubsection{Ownership-Hiding Lawful Contact-Chaining Protocol}

\label{sec-proto2}

The previous version of the protocol allows agencies to learn which telecoms own the phone numbers in its encrypted output. This subsection presents a modification of the previous version of the protocol, which uses a DC-nets-based \emph{anonymity protocol} to hide this information from the agencies (except with respect to the initial target $x$).

An anonymity protocol is run by a number of parties, some of which have messages to send. At the end of the protocol, all parties must learn all messages sent, but no party other than the sender of any given message can learn which party sent that message. Dissent~\cite{dissent} and Verdict~\cite{verdict} both satisfy our requirements; they are more powerful than we need, however, because we assume all telecoms are honest-but-curious.

We can use an anonymity protocol to allow the correct telecom to respond anonymously in steps \ref{proto1:first-send} and \ref{proto1:second-send} in the protocol above. This removes the need for the agencies to know which telecom owns which ciphertext.

Now we can present the following modified protocol. This protocol uses the same data structures as in section~\ref{sec-proto1}, except that $\mathbf{Q}$ now contains pairs $(\Enc_{T(a)}(a), j)$ for unexplored vertices $a$ (omitting the identity of $T(a)$.

\begin{enumerate}

\item The agencies start by agreeing upon a warrant $(x, k, d)$, where $x$ is the target phone number, $k$ is a maximum distance, and $d$ is an upper limit on the degree of vertices to chain through. They encrypt $x$ under the public key of $T(x)$.

\item The agencies initialize a queue $\mathbf{Q}$. Initially, $\mathbf{Q}$ contains only the pair $(\Enc_{T(x)}(x), k)$.

\item The agencies initialize the output list $\mathbf{C}$ to be empty.

\item Each telecom $T$ initializes its output lists $\mathbf{L}_T$ to be empty.

\item While $\mathbf{Q}$ is not empty, do the following:

\begin{enumerate}

\item \label{proto2:dequeue} The agencies dequeue a pair $(\Enc_{T(a)}(a), j)$ from $\mathbf{Q}$. They send the pair $(\Enc_{T(a)}(a), j)$ to all telecoms.

\item All telecoms attempt to decrypt $\Enc_T(a)(a)$ with their decryption keys. Only $T(a)$ will be able to do so. Other telecoms skip to step~\ref{proto2:anon}.

\item $T(a)$ adds $a$ to $\mathbf{L}_T$.

\item \label{proto2:agencycipher} $T(a)$ encrypts $a$ under the agencies' public keys, producing the agency ciphertext $\Enc_\mathcal{A}(a)$.

\item \label{proto2:telecomcipher} If $j>0$, $T(a)$ encrypts each neighbor $b$ of $a$ under the public key of $T(b)$, creating a telecom ciphertext for $b$.

\item \label{proto2:anon} All parties to this protocol engage in the anonymity protocol. $T(a)$ sends an anonymous message consisting of the agency ciphertext it generated in step~\ref{proto2:agencycipher}; the set of telecom ciphertexts generated in step \ref{proto2:telecomcipher}, and $\de(a)$, the number of telecom ciphertexts being sent. The agencies and all telecoms that could not decrypt $\Enc_{T(a)}(A)$ participate but send no anonymous message.

\item When the anonymity protocol is complete, the agencies receive all the ciphertexts. They add $\Enc_\mathcal{A}(a)$ to $\mathbf{C}$.

\item If $\de(a)>d$ and $j\neq k$ (i.e. $a\neq x$), the agencies discard all telecom ciphertexts received for $a$'s neighbors (i.e., agencies refuse to sign these ciphertexts in future steps of the protocol, and do not send them on to the telecoms).

\item Otherwise, for each telecom ciphertext received, the agencies add $(\Enc_{T(b)}(b), j-1)$ to $\mathbf{Q}$.

\end{enumerate}

\item The agencies' final output is the list $\mathbf{C}$. Each telecom $T$'s final output is $\mathbf{L}_T$.

\end{enumerate}

The protocol replaces each query in the protocol of section~\ref{sec-proto1} with broadcast of the telecom ciphertext to all telecoms, and replaces each response with a round of the anonymity protocol. This allows the telecom that owns each phone number to respond with appropriate information about the phone number, but shields the telecom's identity from the agencies (and incidentally from other telecoms).

As in the previous section, It should be noted that the agencies and telecoms need not handle one ciphertext at a time. The agencies can in principle dequeue all of $\mathbf{Q}$ in step \ref{proto2:dequeue} and broadcast all pending vertices to the telecoms. In step \ref{proto2:anon}, multiple telecoms can submit multiple messages to a single run of the anonymity protocol, with only those telecoms which were unable to decrypt any vertices submitting no message. The exact number of messages per instance of the anonymity protocol can be tuned for best efficiency.

\subsection{Discussion of Lawful Contact-Chaining}

\label{sec:discuss}

We now take a moment to discuss the correctness and privacy properties of both variants of our lawful contact-chaining protocol.

\subsubsection{Correctness of Output}

The agencies' outputs from the protocols in Sections~\ref{sec-proto1} and \ref{sec-proto2} will be $\mathbf{C}$. $\mathbf{C}$ will contain agency ciphertexts of all phone numbers at most $k$ phone calls away from $x$, considering only vertices of degree at most $d$. This is the desired output. $C$ reveals nothing to any agencies unless they all provide their decryption keys. To continue the process of lawful investigation, the agencies should combine the output $\mathbf{C}$ with other sets of potential suspects (such as from further runs of this protocol, or from cell tower dumps) in a lawful intersection protocol.

\subsubsection{Privacy}

Both versions of the protocol hide the identities of the chained contacts of $x$. They do allow the agencies to learn the distance from $x$ of each ciphertext in their output, but these ciphertexts cannot be resolved to phone numbers without the cooperation of all agencies.

The protocol of section~\ref{sec-proto1} allows the agencies to learn which telecoms owns which ciphertexts in $\mathbf{C}$. This may be a security concern, since some telecoms are relatively small, specialized, or localized to a particular country or region. If the agencies know that such a such a telecom owns an encrypted phone number, this will not allow them to identify the phone number itself, but might convince the agencies to subject that ciphertext to additional scrutiny, up to the point of decrypting it outside the context of lawful surveillance. This would still require the collusion of all agencies, however. Our revised protocol mitigates this concern. Assuming that the anonymity protocol used in section~\ref{sec-proto2} does not allow its participants to learn who sends each message, then the revised protocol does not leak ciphertext ownership information.

The telecoms learn two types of information as part of the lawful contact chaining protocol. First, they learn the warrant. Second, they learn which of the phone numbers they serve have been captured (in encrypted form) by the protocol, and when they were captured. The telecoms might possibly be able to infer some extra information about $G$ from observing when vertices they own are queried by the agencies, but only of a very limited form. For instance, an agency may serve two phone numbers, $a$ and $b$, which the agencies query at distance 1 and 4 from $x$, respectively . In that case, the telecom can infer that there exists a path in $G$ of length 3 between $a$ and $b$. The telecom does not learn which other phones are involved in that path, and is already aware of all paths of length 2 or less between phone numbers it serves. Therefore, this potential information leak is of little concern.

\subsubsection{Hiding Information From Telecoms}

\label{sec:oblivious}

In both versions of our protocol, the telecoms learn which of their numbers have been submitted to the agencies. They do not know which phone numbers the agencies will actually investigate after running the privacy-preserving set intersection protocol, but they do know which ones \emph{could} be under investigation. Since many numbers \emph{could} be investigated, this does not compromise the agencies' investigative power.

We may point out nevertheless that a modification of our protocol from \ref{sec-proto1} could allow the agencies to hide from the telecoms which of their clients is being surveilled. The telecoms would need to precompute agency ciphertexts for all of their client numbers, and telecom ciphertexts for all of their clients' contacts. With these precomputed databases, the telecoms could then use oblivious transfer to blindly serve the agencies' requests for information about their clients.

\subsection{Performance of Privacy-Preserving Contact Chaining Protocol}

\label{sec:implementation}

We implemented the privacy-preserving contact chaining search protocol of~\ref{sec-proto1} in Java and tested the implementation's running time, CPU time used, and data sent over the network. Below, we describe our implementation and its experimental setup, and then summarize our results.

\subsubsection{Java Implementation}

Our Java implementation uses the variant of our protocol in which the agencies completely exhaust the search queue $\mathbf{Q}$ each round, sending all queries at any given distance from $x$ to the telecoms at once in batches. Ths variant allows for greater parallelism. All of the telecoms receive their batch of queries at the same time, and operate on those queries using eight parallel threads of computation.

We use 2048-bit DSA signatures, 2048-bit RSA encryption for the telecoms, and ElGamal encryption for the agencies' output to provide compatability with the lawful intersection protocol of~\cite{sff-foci2014}.

Our Java program supports any number of agencies and telecoms, but we chose to run tests with three government agencies and four telecoms. Each agency and telecom has a dedicated server in our cloud testbed. As mentioned in~\cite{sff-foci2014}, three is a reasonable choice for the number of agencies, corresponding to three branches of government. Four telecoms should cover most users in any given mobile phone market, and increasing the number of telecoms in our experiments only serves to decrease the protocol's total running time by splitting the same users over more servers.

\subsubsection{Experimental Setup}

For our underlying contact graph, we used an anonymized data set provided by~\cite{snapnets} containing 1.6 million users from Pokec, a Slovakian social network. To replicate the multi-provider environment of the real telephone network, we assigned each user to one of four telecom servers. The telecoms were each given a different number of the users, in proportion to the subscriber base of the largest four telecoms in the world~\cite{mobiforge}.

To experiment with differently sized output sets, we ran our protocol many times, varying $x$, $k$, and $d$. We chose a variety of different-degree starting targets $x$, varied the maximum path length $k$ between 2 and 3, and varied $d$ from 25 to 500. For each run, we measured the total running time of the protocol, the CPU time spent by the agencies and telecoms, and the amount of data sent over the network in total.

These results are important in evaluating how practical our lawful contact-chaining protocol would be it were put into practice by government agencies and telecoms. However, our data set is relatively small compared to the databases held by real telecommunications companies, and each company handles that data using different technologies. The absolute running time and CPU usage of executing this protocol could vary from telecom to telecom. Therefore, we also produced a implementation of the contact-chaining protocol which omits all cryptographic operations. This version of the protocol does not preserve the privacy of users. By comparing the performance of our lawful contact-chaining protocol with the zero-cryptography contact-chaining protocol, however, we can get a sense of the ``cost'' of privacy and accountability as compared to the practice of releasing plaintext data to government surveillance.

\subsubsection{Results}

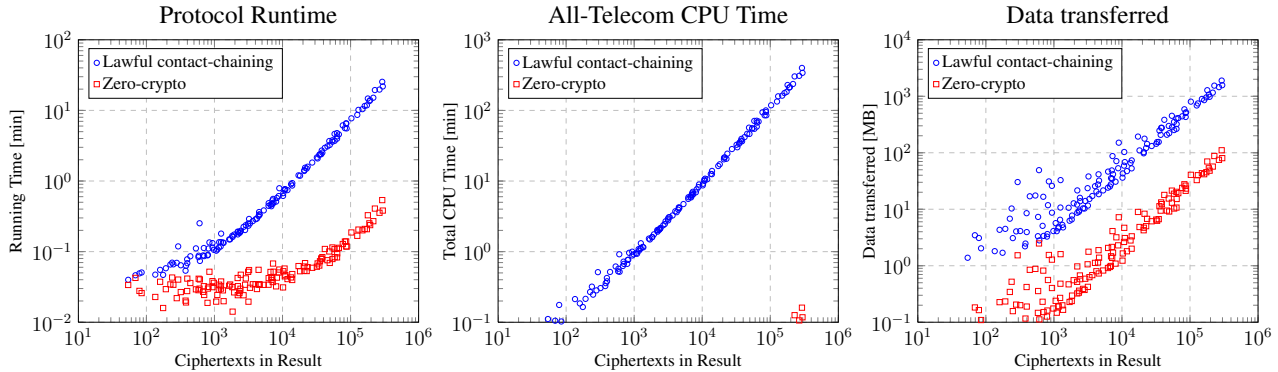
\begin{figure*}[t]
\centering
\begin{subfigure}{0.66\columnwidth}
\centering
\begin{tikzpicture}[scale=0.66]
\begin{loglogaxis}[
	title={Protocol Runtime},
	title style={font=\Large,},
	xlabel={Ciphertexts in Result},
	ylabel={Running Time [min]},
	ylabel style={overlay},
	ticklabel style = {font=\large},
	xmin=10, xmax=1000000,
	ymin=0.01, ymax=100,
	legend pos=north west,
	ymajorgrids=true,
	xmajorgrids=true,
	grid style=dashed,
	legend style={cells={anchor=west}},
]
\addplot[
	color=blue,
	mark=o,
	only marks,
	mark size=1.5pt,
	]
	table [x=Ciphertexts, y=RunningTime, col sep=comma] {chainingrsa.csv};	
\addplot[
	color=red,
	mark=square,
	only marks,
	mark size=1.5pt,
	]
	table [x=Ciphertexts, y=RunningTime, col sep=comma] {chainingnocrypto.csv};
\legend{Lawful contact-chaining, Zero-crypto}
\end{loglogaxis}
\end{tikzpicture}
\label{fig:runtime}
\end{subfigure}\hfill
\begin{subfigure}{0.66\columnwidth}
\centering
\begin{tikzpicture}[scale=0.66]
\begin{loglogaxis}[
	title={All-Telecom CPU Time},
	title style={font=\Large,},
	xlabel={Ciphertexts in Result},
	ylabel={Total CPU Time [min]},
	ylabel style={overlay, anchor=north,},
	ticklabel style = {font=\large},
	xmin=10, xmax=1000000,
	ymin=0.1, ymax=1000,
	legend pos=north west,
	ymajorgrids=true,
	xmajorgrids=true,
	grid style=dashed,
	legend style={cells={anchor=west}},
]
\addplot[
	color=blue,
	mark=o,
	only marks,
	mark size=1.5pt,
	]
	table [x=Ciphertexts, y=CPUTime, col sep=comma] {chainingrsa.csv};
\addplot[
	color=red,
	mark=square,
	only marks,
	mark size=1.5pt,
	]
	table [x=Ciphertexts, y=CPUTime, col sep=comma] {chainingnocrypto.csv};
\legend{Lawful contact-chaining, Zero-crypto}
\end{loglogaxis}
\end{tikzpicture}
\label{fig:cputime}
\end{subfigure}\hfill
\begin{subfigure}{0.66\columnwidth}
\centering
\begin{tikzpicture}[scale=0.66]
\begin{loglogaxis}[
	title={Data transferred},
	title style={font=\Large,},
	xlabel={Ciphertexts in Result},
	ylabel={Data transferred [MB]},
	ylabel style={overlay, anchor=north,},
	ticklabel style = {font=\large},
	xmin=10, xmax=1000000,
	ymin=0.1, ymax=10000,
	legend pos=north west,
	ymajorgrids=true,
	xmajorgrids=true,
	grid style=dashed,
	legend style={cells={anchor=west}},
]
\addplot[
	color=blue,
	mark=o,
	only marks,
	mark size=1.5pt,
	]
	table [x=Ciphertexts, y=Bytes, col sep=comma] {chainingrsa.csv};	
\addplot[
	color=red,
	mark=square,
	only marks,
	mark size=1.5pt,
	]
	table [x=Ciphertexts, y=Bytes, col sep=comma] {chainingnocrypto.csv};
\legend{Lawful contact-chaining, Zero-crypto}
\end{loglogaxis}
\end{tikzpicture}
\label{fig:data}
\end{subfigure}
\captionsetup{justification=centering}
\caption{Performance of Lawful Contact-Chaining}
\label{fig:performance}
\end{figure*}

Our implementation of lawful contact-chaining performed well. Our experiments showed a linear relationship between the number of ciphertexts in the output and the running time, CPU time, and data usage of the protocol. We display graphs of our recorded data in Figure~\ref{fig:performance}. Taking the average of all cases with $d>25$, the telecoms used 58.2 ms of CPU time per ciphertext. The agencies used, again in the average case, 2.0 ms of CPU time per ciphertext. Note that these times are the sums taken over all telecoms and all agencies respectively. Because the agencies have do very little cryptography in this protocol, we focus on the telecoms' CPU time in our evaluation. 

We found that our protocol was able to process, in the average case, 197.4 ciphertexts per second. To return to our example from earlier of a network with an average of 30 contacts per user, a lawful contact-chaining search with $k=2$ would have 900 users in the output, and a search with $k=3$ would have 27,000 users in the output. To compare these times to some of our acutal experiments, we found that a search that returned 937 ciphertexts took 6.86 seconds to run, and a search that returned 27,338 ciphertexts took 109.55 seconds to run. To provide another point of comparison, Bandits~\cite{sff-foci2014} refers to the ``High Country Bandits'' case, in which the FBI performed an intersection of 150,000 phone number to help solve a series of bank robberies. In one of our experiments with lawful contact chaining, we find that a similarly sized data set of 149,535 ciphertexts took 625.08 seconds - 10.4 minutes - to compile with our protocol. Given the context of a criminal investigation, we feel these running times are quite reasonable. 

The zero-cryptography version of our program ran, predictably, more quickly than the lawful privacy-preserving version. The total CPU time across all telecoms needed for our zero-crypto implementation never rose above ten seconds, even in the largest cases. This result allows us to disambiguate the cost of \emph{information retrieval} from \emph{privacy protection}. The linear relationship between the size of the encrypted user data set and the performance in terms of running time, CPU time, and network data usage of the protocol all remain even when we subrtract out the time to run all non-cryptgraphic parts of the protocol. We therefore conclude that, even given the potental database operations real telecoms would have to contend with, the cost of adding privacy-preservation to the contact-chaining protocol will remain reasonable.

%% file: related.tex
\section{Related Work}\label{sec-related}
Privacy-preserving computation has been studied extensively. An overview
of the approaches taken by the cryptographic-research community is provided by
Perry {\it et al.}~\cite{pgfw-scn2014}; an overview of of the data-mining
approach is provided by Aggarwal and Yu~\cite{ay2008}.  

Kamara~\cite{kamara}
and Kroll {\it et al.}~\cite{kroll} used cryptographic protocols to 
achieve privacy and accountability in the surveillance of known targets.
Bandits~\cite{sff-foci2014} formulated the openness principle that
we have followed and were the first to design privacy-preserving protocols for
the surveillance of unknown targets.  

Kearns {\it et al.}~\cite{krwy-pnas16} present efficient graph-search 
algorithms that distinguish targeted users from untargeted users; for each 
untargeted user $u$, the set of direct contacts of $u$ remains private.
Unlike our privacy-preserving contact-chaining algorithms, which rely
on cryptographic techniques, their graph-search algorithms rely on 
differential-privacy techniques.

%% file: conc.tex
\section{Open Problems and Future Work}\label{sec-future}
Bandits noted in \cite[Section 6.1]{sff-foci2014} that
set intersection is but one type of computation that can be of use
to law-enforcement and intelligence agencies.  They observed that it 
would be interesting to identify other such computational problems 
and to devise accountable, privacy-preserving protocols to solve them.  
The work in this paper on contact chaining represents progress in that
direction.  

Another problem of potential interest is the retrieval of
targeted users' postings on Facebook and other social networks, including 
those that are shared only with a small subset of the targeted user's 
``friends.'' Accountable surveillance of social-network postings may present 
novel protocol-design challenges, because it deals with one-to-many 
communication, whereas previous work in the area dealt with pairwise 
communication.

For contact chaining, it may be possible to speed up our protocols by using
elliptic-curve cryptography instead of RSA. Additionally, our assumption that
all parties behave in an honest-but-curious manner might be weakened. By using
standard zero-knowledge proof techniques, it might be possible to create versions
of the protocols in Section~\ref{sec-chaining} that are secure against, for
example, a rogue agent's maliciously modifying telecom-supplied data in order
to falsely incriminate a victim. It may also be interesting to 
generalize the differential-privacy approach of Kearns {\it et 
al.}~\cite{krwy-pnas16} so that it applies to indirect contacts as well as
direct contacts.

Finally, the Openness Principle put forth in \cite{sff-foci2014} is but one step
toward a full understanding of how democratic processes and the rule of law
can be carried into the digital world.  Further investigation, much of it
interdisciplinary, is needed.

\subsection*{Acknowledgments}

The first author was supported by a Google Faculty Research Award.
The second author was supported by NSF grants CNS-1407454 and CNS-1409599, DHS grant FA8750-16-2-0034, and William and Flora Hewlett Foundation grant 2016-3834.
The third author was supported by EPFL, the AXA Research Fund, and DHS grant FA8750-16-2-0034.